\documentstyle[aps,multicol,epsfig]{revtex}
\begin{document}
\draft
\title{Tunneling effect on composite fermion pairing state in bilayer
quantum Hall system}
\author{Takao Morinari}
\address{Yukawa Institute for Theoretical Physics,
Kyoto University, Kyoto 606-8502, Japan}

\date{\today}
\maketitle
\begin{abstract}
We discuss the composite fermion pairing state in bilayer quantum Hall 
systems.
After the evaluation of the range of the inter-layer
separation in which the quantum Hall state is stabilized,
we discuss the effect of inter-layer tunneling on the composite
fermion pairing state at $\nu=1/2$.
We show that there is a cusp at the transition point between the
Halperin $(3,3,1)$ state and the Pfaffian state.
\end{abstract}


\begin{multicols}{2}
\narrowtext
\section{Introduction}
As a remarkable example of strongly correlated electron systems,
the fractional quantum Hall effect has been studied
extensively.\cite{review,review2}
In the fractional quantum Hall systems, the Coulomb interaction plays
a dominant role.
Due to the presence of the strong magnetic field, the Coulomb
interaction gives rise to a two-body correlation of non-zero relative
angular momentum.
The Laughlin wave function,\cite{LAUGHLIN} which captures
the essential properties of the system, consists of this two-body
correlation only.
The role of the Coulomb interaction to generate the two-body
correlation becomes clear when it is described in terms of Haldane's
pseudopotential.\cite{HALDANE}
From the analysis of the pseudopotentials, one can see that
the most fundamental contribution comes from the short-range Coulomb
interaction.

In the single-layer quantum Hall systems, it is hard to change the
short-range Coulomb interaction.
Whereas in the bilayer quantum Hall systems, we can adjust the
inter-layer Coulomb interaction by varying the inter-layer separation
$d$.
In addition, there is another parameter, the inter-layer tunneling
amplitude $t$ to control the system.
Since there are such additional parameters to control the system, we
can expect rich phase diagram in the bilayer quantum Hall systems.

In the absence of inter-layer tunneling, the properties of the system
seems to be well described by the Halperin $(m,m,n)$ wave
function\cite{HALPERIN_MMN}, which is an extention of the Laughlin
wave function:
\begin{eqnarray}
\lefteqn{\Psi_{mmn} (z_1^{\uparrow},z_2^{\uparrow},\cdots,z_N^{\uparrow};
z_1^{\downarrow},z_2^{\downarrow},\cdots,z_N^{\downarrow})} 
\nonumber \\
&=&\prod_{i<j} (z_i^{\uparrow}-z_j^{\uparrow})^m
\prod_{i<j} (z_i^{\downarrow}-z_j^{\downarrow})^m
\prod_{i,j} (z_i^{\uparrow}-z_j^{\downarrow})^n 
\nonumber \\ & & \times 
\exp \left[ - \frac{1}{4\ell_B^2} 
\sum_j (|z_j^{\uparrow}|^2 + |z_j^{\downarrow}|^2) \right],
\label{mmn_wf}
\end{eqnarray}
where $\uparrow$ and $\downarrow$ are the indices for layers and
$z_j^{\sigma} = x_j^{\sigma} - i y_j^{\sigma}$ is the complex
coordinate of the $j$-th electron.
The integer numbers $m$ and $n$ is associated with 
the total Landau level filling $\nu$ as $\nu=2/(m+n)$.

At $\nu=1/2$, the $(3,3,1)$ wave function has good overlap with the
wave function of the finite size system \cite{YOSHIOKA_ETAL}
in the absence of the inter-layer tunneling and in the appropriate
region of the inter-layer separation.
It was pointed out by Ho\cite{HO} that the $(3,3,1)$ state and
the Pfaffian state \cite{MOORE_READ}, which 
may be stabilized in the strong inter-layer tunneling
limit,\cite{HALPERIN_BI}
are unified to the form of the p-wave pairing state of composite
fermions.

In this paper, we discuss the stability of the composite fermion
paring states in the bilayer quantum Hall systems.
First, we discuss the range of the inter-layer separation $d$ in which
the $(m,m,n)$ state is stabilized and the appropriate choice of the
number of fluxes for composite fermions.
After deriving the mean field equations for the composite fermion
pairing states,
we consider the effect of the inter-layer tunneling $t$ on the p-wave
pairing state at $\nu=1/2$.
The evolution of the $(3,3,1)$ state to the Pfaffian state is shown
being based on the analysis of the gap equation.
We also examine the possibility of Haldane-Rezayi
state\cite{HALDANE_REZAYI} at $\nu=1/2$.

The outline of this paper is as follows.
In Sec.~\ref{sec_two}, we determine the appropriate choice of $(m,n)$
and we describe the relationship between the Halperin $(m,m,n)$ state
and the p-wave pairing state of composite fermions.
The formulation for the analysis of the pairing state is given in
Sec.~\ref{sec_cfp}.
The gap equations for the triplet pairing state and the singlet
pairing state are derived in the presence of the inter-layer
tunneling.
In Sec.~\ref{sec_t}, we discuss the inter-layer tunneling effect.
Finally, we summarize the results in Sec.~\ref{sec_summary}.

\section{Two-body correlation}
\label{sec_two}
In this section, we discuss the bilayer quantum Hall systems in the
absence of the inter-layer tunneling in order to find the appropriate
choice of the number of attached fluxes for composite fermions.
For the determination of those numbers, the effect of inter-layer
tunneling may be negligible because they are associated with the
two-body correlation due to the short-range Coulomb interaction as
discussed in Introduction.

Since the two-body correlations, which is connected with 
the numbers $m$ and $n$ in the $(m,m,n)$ wave function
(\ref{mmn_wf}), is associated with the short-range Coulomb
interaction, $m$ and $n$ are determined by ``high-enrgy'' physics.
We compare Haldane's pseudopotentials with various choices of $m$
and $n$ \cite{MORINARI_BI}
because ``high-energy'' physics is governed by Haldane's
pseudopotentials.

The basis for the two-body electron correlations is given by the wave
function for an electron pair with the relative angular momentum $m$
and the angular momentum of the central motion being zero:
\begin{eqnarray}
\langle z_1,z_2|\psi_m \rangle
&=&
\frac{1}{\sqrt{4^{m+1}\ell_B^{2m+4}\pi^2 m!}}
\left( z_1-z_2 \right)^m  
\nonumber \\ & & \times 
\exp \left[ -\frac{1}{4\ell_B^2}
\left( |z_1|^2+|z_2|^2 \right) \right],
\label{eq_2}
\end{eqnarray}
where $\ell_B = \sqrt{c\hbar/eB}$ is the magnetic length.
Since Haldane's pseudopotential is equivalent to the Coulomb energy 
estimated in first order, the total energy for
``high-energy'' physics for total $N$ electrons is given by
\begin{equation}
E_C^{(2)}(m,n)
= \frac{N(N-1)}{2} \times \epsilon (m,d=0) \times 2
+ N^2 \times \epsilon (n,d),
\label{pre_Ec2}
\end{equation}
where 
\begin{eqnarray}
\epsilon (m,d) 
&=& \langle \psi_m| \frac{e^2}{\epsilon \sqrt{r^2+d^2}}|\psi_m \rangle
\nonumber \\ &=& 
\frac{e^2}{\epsilon \ell_B} \times \frac{1}{m!}
\int_0^{\infty} dx \frac{x^{2m+1}}{\sqrt{x^2+\lambda^2}} {\rm e}^{-x^2},
\end{eqnarray}
with $\epsilon$ the dielectric constant and $\lambda
=d/2\ell_B$. 
In the thermodynamic limit, we obtain
\begin{equation}
E_C^{(2)}(m,n)/N^2=\epsilon (m,d=0) + \epsilon (n,d).
\label{eq_e2}
\end{equation}
Although the two-body correlation energies are manifestly
overestimated in $E_C^{(2)}(m,n)$, the right hand side of
Eq.~(\ref{eq_e2}) may be reliable as far as we are concerned with 
the pseudopotentials.
For the choice of $m$ and $n$, we cannot choose arbitrary pair of
$m$ and $n$. 
There is a constraint on the choice of $m$ and $n$.
From the Halperin $(m,m,n)$ wave function,
one can see that the angular momentum of the electron at the edge of the 
sample is equal to $(N-1) \times m+N\times n \equiv M$.
Since the wave function of this electron is proportional to 
$z^M {\rm e}^{-r^2/4\ell_B^2}$, the density of it has its maximum
at $r=\sqrt{2M} \ell_B \equiv R$.
Of course $\pi R^2$ is the area of the system.
Taking the thermodynamic limit $N \rightarrow \infty$, we obtain
\begin{equation}
2\pi \ell_B^2 \times N \times (m+n) = \Omega.
\label{mn_area}
\end{equation}
We consider the case of symmetric electron density
$\rho_{\uparrow}=\rho_{\downarrow}$.
In this case, Eq.~(\ref{mn_area}) is reduced to 
\begin{equation}
m + n = 2/\nu.
\label{constraint}
\end{equation}

Now we determine $m$ and $n$ which gives the lowest $E_C^{(2)}(m,n)$
under the constraint (\ref{constraint}).
For the case of $\nu=1/2$, the constraint (\ref{constraint})
is $m+n=4$. Therefore, the possible choice of $(m,n)$ is,
$(4,0)$, $(3,1)$ and $(2,2)$.
The pair $(m,n)$ with $m<n$ always has larger energy than that with
$m\geq n$.
Note that in case of $(m,n)=({\rm even},{\rm even})$, the system is
not the quantum Hall state.
It is compressible liquid of composite fermions and the total wave
function can not be determined by the two-body correlations only.

\begin{figure}
\center
\epsfxsize=2.0truein
\psfig{file=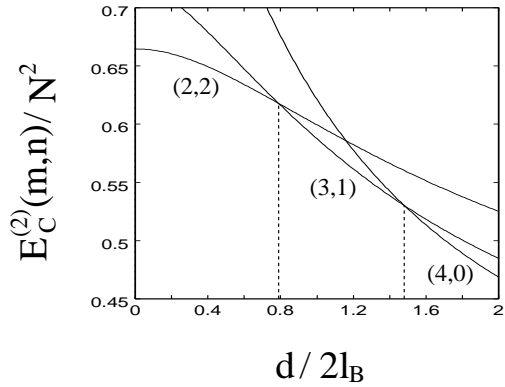,width=2.0in,angle=270}
\vspace{0.1in}
\caption{The two-body correlation energy $E_C^{(2)}(m,n)/N^2$ for
$(m,n)=(4,0),(3,1)$, and $(2,2)$ in units of $e^2/\epsilon \ell_B$.}
\label{nu_half}
\end{figure}

In Fig.~\ref{nu_half}, we show the energy $E_C^{(2)} (m,n)/N^2$ for 
$(m,n)=(4,0),(3,1)$, and $(2,2)$.
The region where the choice of $(m,n)=(3,1)$ 
gives the lowest energy $E_C^{(2)}(m,n)$ is $0.789 < d/2\ell_B < 1.480$.
The Halperin $(3,3,1)$ state is stabilized in this region.

\begin{figure}[htbp]
\center
\epsfxsize=2.0truein
\psfig{file=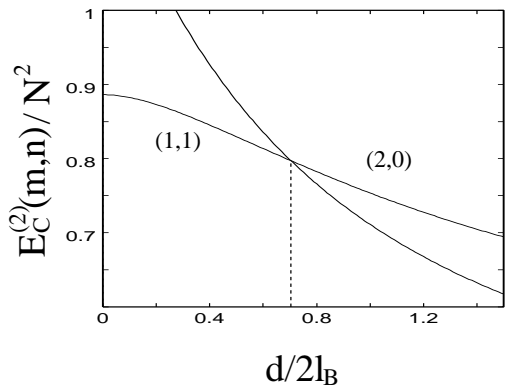,width=2.0in,angle=270}
\vspace{0.1in}
\caption{The energy $E_C^{(2)}(m,n)/N^2$ for $(m,n)=(2,0)$, and
$(1,1)$
in units of $e^2/\epsilon \ell_B$.}
\label{nu_one}
\end{figure}

For the case of $\nu=1$, the constraint (\ref{constraint})
is $m+n=2$. Therefore, the possible choice for $(m,n)$ is
$(2,0)$ and $(1,1)$.
In Fig.\ref{nu_one}, we show the energy $E_C^{(2)}(m,n)/N^2$ for
$(2,0)$ and $(1,1)$.
The region where the choice of $(m,n)=(1,1)$ gives the lowest energy
$E_C^{(2)}(m,n)$ is $0 < d/2\ell_B < 0.703(\equiv d_c)$.
The Halperin $(1,1,1)$ state is stabilized in this region.
Though the above estimation is crude, the critical value $d_c$ for
$\nu=1$ is close to $2 \ell_B$ that was obtained by 
Murphy {\it et al.} experimentally \cite{MURPHY}.

In general, the estimation of $E^{(2)}_C (m,n)$ shows that 
the pair $(m,n)$ giving the lowest $E^{(2)}_C (m,n)$ is $(2/\nu,0)$ 
for $d \gg \ell_B$. As we decrease the value of $d$, it
changes as $(2/\nu,0) \rightarrow (2/\nu-1,1) \rightarrow \cdots
\rightarrow (1/\nu,1/\nu)$.
In this sequence, the quantum Hall state is stable at $(m,n)=({\rm
odd},{\rm odd})$, whereas the compressible state of composite fermions 
is stable at $(m,n)=({\rm even},{\rm even})$.

Now we discuss the relationship between the $(m,m,n)$ wave function
and the p-wave pairing state of composite fermions.\cite{HO}
As shown in Ref.~\cite{HO}, the wave function of the p-wave pairing
state of composite fermions at $\nu=1/2$ is, in the second quantized
form,
\begin{eqnarray}
|N,\chi \rangle 
&=&
\int \prod_{j=1}^{N} d^2 z_j 
\prod_{i<j} (z_i-z_j)^2
{\rm e}^{-\frac{1}{4\ell_B^2} \sum_{j=1}^{N} |z_j|^2} 
\nonumber \\ & & \times
\prod_{j=1}^{N} \left[ \frac{\sum_{\sigma \sigma'}
\chi_{\sigma \sigma'} 
\psi_{\sigma}^{\dagger} (z_{2j-1}) \psi_{\sigma'}^{\dagger} (z_{2j})}
{z_{2j-1}-z_{2j}} \right]~|0\rangle,
\end{eqnarray}
where $\psi_{\sigma}^{\dagger} (z)$ is the creation operator of the
electron at $z$ with spin $\sigma$.
For the $(3,3,1)$ state, $\chi$ is given by
$\chi= \left[\begin{array}{cc} 0 & 1 \\ 1 & 0 \end{array} \right]$.
Note that in this case the p-wave pairing wave function is the
$(1,1,-1)$ state.\cite{HALDANE_REZAYI,HO}
Therefore, in general the $(m,m,n)$ state is described by the p-wave 
pairing state with $\chi= \left[\begin{array}{cc} 0 & 1 \\ 1 & 0
\end{array} \right]$ of composite fermions with the number of attached
fluxes $(\phi_1,\phi_2)=(m-1,n+1)$.
(The even integer $\phi_1$ ($\phi_2$) is for the intra(inter)-layer
correlations.)
For the Pfaffian state, 
$\chi$ is given by $\chi=\left[\begin{array}{cc} 1/\sqrt{2} &
1/\sqrt{2} \\ 
1/\sqrt{2} & 1/\sqrt{2} \end{array} \right]$.
If the system is not the quantum Hall state, then the wave function of 
composite fermions is not the form of the pairing state.

\section{Composite fermion pairing}
\label{sec_cfp}
In the last section, we have discussed the appropritate choice of the 
number of attached fluxes for composite fermions.
Now we introduce composite fermions in the second quantized form:
\begin{equation}
\left\{ 
 \begin{array}{l}
   \tilde{\psi}_{\alpha} ({\bf r}) 
= {\rm e}^{-iJ_{\alpha} ({\bf r})} \psi_{\alpha} ({\bf r}), \\
   \tilde{\psi}^{\dagger}_{\alpha} ({\bf r}) 
= \psi_{\alpha}^{\dagger} ({\bf r}) {\rm e}^{iJ_{\alpha} 
({\bf r})}, 
 \end{array}
\right.
\label{eq_cf}
\end{equation}
where the function $J_{\alpha} ({\bf r})$ is given by
\begin{equation}
J_{\alpha} ({\bf r}) = \sum_{\beta} K_{\alpha \beta}
 \int d^2 {\bf r}^{\prime} \rho_{\beta} \left( {\bf r}^{\prime} \right)
 {\rm Im} \ln \left( z - z^{\prime} \right).
\label{eq_J}
\end{equation}
Here $\rho_{\alpha} ({\bf r}) 
= \psi^{\dagger}_{\alpha} ({\bf r}) \psi_{\alpha} ({\bf r})
= \tilde{\psi}^{\dagger}_{\alpha} ({\bf r}) 
\tilde{\psi}_{\alpha} ({\bf r})$ and
$K=\left( \begin{array}{cc} \phi_1 & \phi_2 \\ \phi_2 & \phi_1
\end{array} \right)$ with $\phi_1$ and $\phi_2$ being even integer.
In terms of the composite fermion fields $\tilde{\psi}^{\dagger}$ 
and $\tilde{\psi}$, the kinetic energy term of the electrons is
rewritten as
\begin{eqnarray}
H_0 
&=& 
\sum_{\alpha} \frac{1}{2m} \int d^2 {\bf r}
\psi^{\dagger}_{\alpha} ({\bf r}) 
\left( -i \hbar \nabla + \frac{e}{c} {\bf A} \right)^2
\psi_{\alpha} ({\bf r}) \nonumber \\
&=& 
\sum_{\alpha} \frac{1}{2m} \int d^2 {\bf r}
\tilde{\psi}^{\dagger}_{\alpha} ({\bf r}) 
\left( -i \hbar \nabla + \frac{e}{c} {\bf A} 
+ \frac{e}{c} {\bf a}_{\alpha}
\right)^2
\tilde{\psi}_{\alpha} ({\bf r}).\nonumber \\
\label{eq_H0}
\end{eqnarray}
Here ${\bf a}_{\alpha}$ is the Chern-Simons gauge field,
\begin{equation}
{\bf a}_{\alpha} ({\bf r}) = - \frac{ic\hbar}{e} \nabla
J_{\alpha} ({\bf r}).
\end{equation}
The Chern-Simons gauge field ${\bf a}_{\alpha}$ obeys the constraint
\begin{equation}
\nabla \times {\bf a}_{\alpha} ({\bf r} )
= \phi_0 \sum_{\beta} K_{\alpha \beta} \rho_{\beta} ({\bf r}),
\label{eq_flux}
\end{equation}
where $\phi_0 = ch/|e|$ is the flux quantum.

The first order term of Eq.~(\ref{eq_H0}) with respect to the
fluctuation of the Chern-Simons gauge field 
${\bf A} + {\bf a}_{\alpha}$ yields the minimal coupling term.
Eliminating the Chern-Simons gauge field fluctuations upon using the
constraint (\ref{eq_flux}), we obtain
\begin{eqnarray}
V 
&=& \frac{1}{2\Omega} \sum_{{\bf k}_1 \neq {\bf k}_2, {\bf q}}
\sum_{\alpha \beta} K_{\alpha \beta} 
V_{{\bf k}_1, {\bf k}_2} 
\tilde{\psi}^{\dagger}_{{\bf k}_1+{\bf q}/2,\alpha} 
\tilde{\psi}^{\dagger}_{-{\bf k}_1+{\bf q}/2,\beta} 
\nonumber \\ & & \times 
\tilde{\psi}_{-{\bf k}_2+{\bf q}/2,\beta} 
\tilde{\psi}_{{\bf k}_2+{\bf q}/2,\alpha},
\label{eq_Vp}
\end{eqnarray}
where $V_{{\bf k}_1 {\bf k}_2}=
\frac{4\pi i}{m} {\tilde \phi} \frac{{\bf
k}_1 \times {\bf k}_2}{|{\bf k}_1-{\bf k}_2|^2}$.
This interaction gives rise to an attractive interaction that leads to 
the p-wave pairing state.\cite{GREITER_ETAL}
The second order term of Eq.~(\ref{eq_H0}) with respect to the
fluctuations of the Chern-Simons gauge field yields the three-body
interaction term after eliminating the Chern-Simons gauge field
fluctuations.
From the analysis of non-unitary transformation, this three-body
interaction term turns out to be the counter term to the short-range
Coulomb interaction.\cite{MORINARI}
However, if we restrict ourselves to the range of the inter-layer
separation, $d$ where the states 
based on composite fermions are stabilized, we may neglect the
three-body interaction term.
In addition, we neglect the long-range Coulomb
interaction, which gives rise to a pair-breaking effect,
because the pairing state of compoisite fermions may be
stable in the region where the Halperin $(m,m,n)$ state is stable.
In the following analysis, we concentrate on the analysis of the
pairing interaction and the inter-layer tunneling.

Including the inter-layer tunneling effect, 
$H_t = -t \int d^2 {\bf r} \left[
\psi^{\dagger}_{\uparrow} ({\bf r})
\psi_{\downarrow} ({\bf r})
+
\psi^{\dagger}_{\downarrow} ({\bf r})
\psi_{\uparrow} ({\bf r})
\right]$,
the Hamiltonian for
composite fermions may be written as
\begin{eqnarray}
H &=& \sum_{{\bf k} \alpha \beta} \xi^{{\bf k}}_{\alpha \beta}
 \tilde{\psi}^{\dagger}_{{\bf k} \alpha} \tilde{\psi}_{{\bf k} \beta}
\nonumber \\ & & 
+ \frac{1}{2\Omega} \sum_{{\bf k}_1 \neq {\bf k}_2} \sum_{\alpha \beta}
V_{{\bf k}_1,{\bf k}_2}^{\alpha \beta}
    \tilde{\psi}^{\dagger}_{{\bf k}_1 \alpha} 
\tilde{\psi}^{\dagger}_{-{\bf k}_1 \beta} 
    \tilde{\psi}_{-{\bf k}_2 \beta} \tilde{\psi}_{{\bf k}_2 \alpha},
\label{h_bi}
\end{eqnarray}
where $\xi^{\bf k}_{\uparrow \uparrow} = 
\xi^{\bf k}_{\downarrow \downarrow} = k^2/2m - \mu$ and
$\xi^{\bf k}_{\uparrow \downarrow} 
= \xi^{\bf k}_{\downarrow \uparrow} = -t$.
In the interaction term, we have restricted ourselves to the
scattering processes of pairs with zero total momentum.
Note that the formulation in this section can be easily extended to
the multicomponet systems.

From Eq.(\ref{h_bi}), we can define the mean field Hamiltonian as
\begin{eqnarray}
H_{\rm MF}&=& \frac12
\sum_{\bf k} 
\left(
\begin{array}{cccc}
\tilde{\psi}_{{\bf k}\uparrow}^{\dagger} &
\tilde{\psi}_{{\bf k}\downarrow}^{\dagger} &
\tilde{\psi}_{-{\bf k}\uparrow} &
\tilde{\psi}_{-{\bf k}\downarrow}
\end{array}
\right) 
\nonumber \\ & & \times
\left( 
\begin{array}{cc}
\xi^k & \Delta^{\bf k} \\
\left( \Delta^{\bf k} \right)^{\dagger} & -\xi^k
\end{array}
\right)
\left(
\begin{array}{c}
\tilde{\psi}_{{\bf k}\uparrow} \\
\tilde{\psi}_{{\bf k}\downarrow} \\
\tilde{\psi}_{-{\bf k}\uparrow}^{\dagger} \\
\tilde{\psi}_{-{\bf k}\downarrow}^{\dagger}
\end{array}
\right),
\end{eqnarray}
where the pairing matrix is defined as
\begin{equation}
\Delta^{{\bf k}}_{\alpha \beta} 
= - \frac{1}{2\Omega}  
\sum_{{\bf k} ( \neq {\bf k}^{\prime} )} 
V^{\alpha \beta}_{{\bf k} {\bf k}^{\prime}}
\langle 
\tilde{\psi}_{-{\bf k}^{\prime} \beta} 
\tilde{\psi}_{{\bf k}^{\prime} \alpha}
\rangle.
\label{eq_Delta}
\end{equation}

First we consider the triplet pairing case.
Since we consider the symmetric bilayer systems, we take the symmetric 
form of the pairing matrix:
$\Delta^{{\bf k}}_{\uparrow \downarrow} = 
 \Delta^{{\bf k}}_{\downarrow \uparrow}$ 
and
$\Delta^{{\bf k}}_{\uparrow \uparrow} = 
 \Delta^{{\bf k}}_{\downarrow \downarrow}$.
Diagonalization of the mean field Hamiltonian yields the following gap 
equations at zero temperature:
\begin{equation}
\Delta^{\bf k}_{\uparrow \uparrow}
= - \frac{1}{4\Omega} \sum_{{\bf k}^{\prime} (\neq {\bf k})}
 V_{{\bf k} {\bf k}^{\prime}}^{\uparrow \uparrow} 
\left[ \frac{\Delta^{{\bf k}^{\prime}}_{\uparrow \uparrow} 
+ \Delta^{{\bf k}^{\prime}}_{\uparrow \downarrow}}
{E_{{\bf k}^{\prime}}^+} 
+
\frac{\Delta^{{\bf k}^{\prime}}_{\uparrow \uparrow} 
- \Delta^{{\bf k}^{\prime}}_{\uparrow \downarrow}}
{E_{{\bf k}^{\prime}}^-} \right],
\label{t_gap1}
\end{equation}
\begin{equation}
\Delta^{\bf k}_{\uparrow \downarrow}
= - \frac{1}{4\Omega} \sum_{{\bf k}^{\prime} (\neq {\bf k})}
 V_{{\bf k} {\bf k}^{\prime}}^{\uparrow \downarrow} 
\left[ \frac{\Delta^{{\bf k}^{\prime}}_{\uparrow \uparrow} 
+ \Delta^{{\bf k}^{\prime}}_{\uparrow \downarrow}}
{E_{{\bf k}^{\prime}}^+} 
-
\frac{\Delta^{{\bf k}^{\prime}}_{\uparrow \uparrow} 
- \Delta^{{\bf k}^{\prime}}_{\uparrow \downarrow}}
{E_{{\bf k}^{\prime}}^-} \right],
\label{t_gap2}
\end{equation}
where $E^{\pm}_{\bf k}=\sqrt{(\xi_k-t)^2+|\Delta^{\bf k}_{\uparrow
\uparrow} \pm \Delta^{\bf k}_{\uparrow \downarrow}|^2}$.
Meanwhile for the singlet pairing state, $\Delta^{\bf k}_{\uparrow
\downarrow} = - \Delta^{\bf k}_{\downarrow \uparrow} \equiv
\Delta_{\bf k}$,
the gap equation is given by
\begin{eqnarray}
\Delta_{{\bf k}} &=& - \frac{1}{4\Omega} 
\sum_{{\bf k}^{\prime} (\neq {\bf k})}
V_{{\bf k} {\bf k}^{\prime}}^{\uparrow \downarrow} 
\frac{\Delta_{{\bf k}^{\prime}}}{E_{{\bf k}^{\prime}}}
\left[ \tanh \frac{\beta (E_{{\bf k}^{\prime}}-t)}{2}
\right. \nonumber \\
& & \left. ~~~~~+ 
\tanh \frac{\beta (E_{{\bf k}^{\prime}} + t)}{2} \right],
\end{eqnarray}
where $E_{\bf k}=\sqrt{\xi_k^2 + |\Delta_{\bf k}|^2}$.
At zero temperature, this equation is reduced to 
\begin{equation}
\Delta_{{\bf k}} = - \frac{1}{2\Omega} 
\sum_{{\bf k}^{\prime} (\neq {\bf k}), E_{{\bf k}^{\prime}}>t}
V_{{\bf k}^{\prime} {\bf k}}^{\uparrow \downarrow} 
\frac{\Delta_{{\bf k}^{\prime}}}{E_{{\bf k}^{\prime}}}.
\end{equation}
Note that there is the constraint $E_{{\bf k}^{\prime}}>t$
in the summation over ${\bf k}'$-space.

In the absence of the inter-layer tunneling, we can take 
$\Delta^{\bf k}_{\uparrow \uparrow} =0$ because a pairing state 
with $\Delta^{\bf k}_{\uparrow \uparrow} \neq 0$ 
may be stable only in the vicinity of the sample boundary.
By this choice of the pairing matrix, the gap equation has the same
form both for the triplet pairing state and for the singlet pairing
state:
\begin{equation}
\Delta_{{\bf k}} = - \frac{1}{2\Omega} 
\sum_{{\bf k}^{\prime} (\neq {\bf k})}
V_{{\bf k} {\bf k}^{\prime}}^{\uparrow \downarrow} 
\frac{\Delta_{{\bf k}^{\prime}}}{E_{{\bf k}^{\prime}}}.
\end{equation}
We can solve this gap equation by taking the form of the gap as
$\Delta_{\bf k}=\Delta \epsilon_F f(k,k_F) \exp (-i \ell \theta_{\bf
k})$, where $f(k,k_F)=(k/k_F)^{\ell}$ for $k<k_F$ and 
$f(k,k_F)=(k_F/k)^{\ell}$ for $k>k_F$ and $\ell$ is an
integer.\cite{GREITER_ETAL,MORINARI}
From the analysis of this gap equation, we find that the ground state
is the p-wave pairing state.
Note that the $s$-wave pairing state is excluded
because the pairing interaction originates from the Lorentz
interaction induced by the Chern-Simons gauge field
interaction.\cite{MORINARI}
At $\nu=1/2$, the p-wave pairing state corresponds to the $(3,3,1)$
state.
At $\nu=1$ with the choice of $(\phi_1,\phi_2)=(0,2)$,
the p-wave state corresponds to the $(1,1,1)$ state.\cite{MORINARI_BI}

\section{Effect of inter-layer tunneling}
\label{sec_t}
\begin{figure}[htbp]
\center
\epsfxsize=2.0truein
\psfig{file=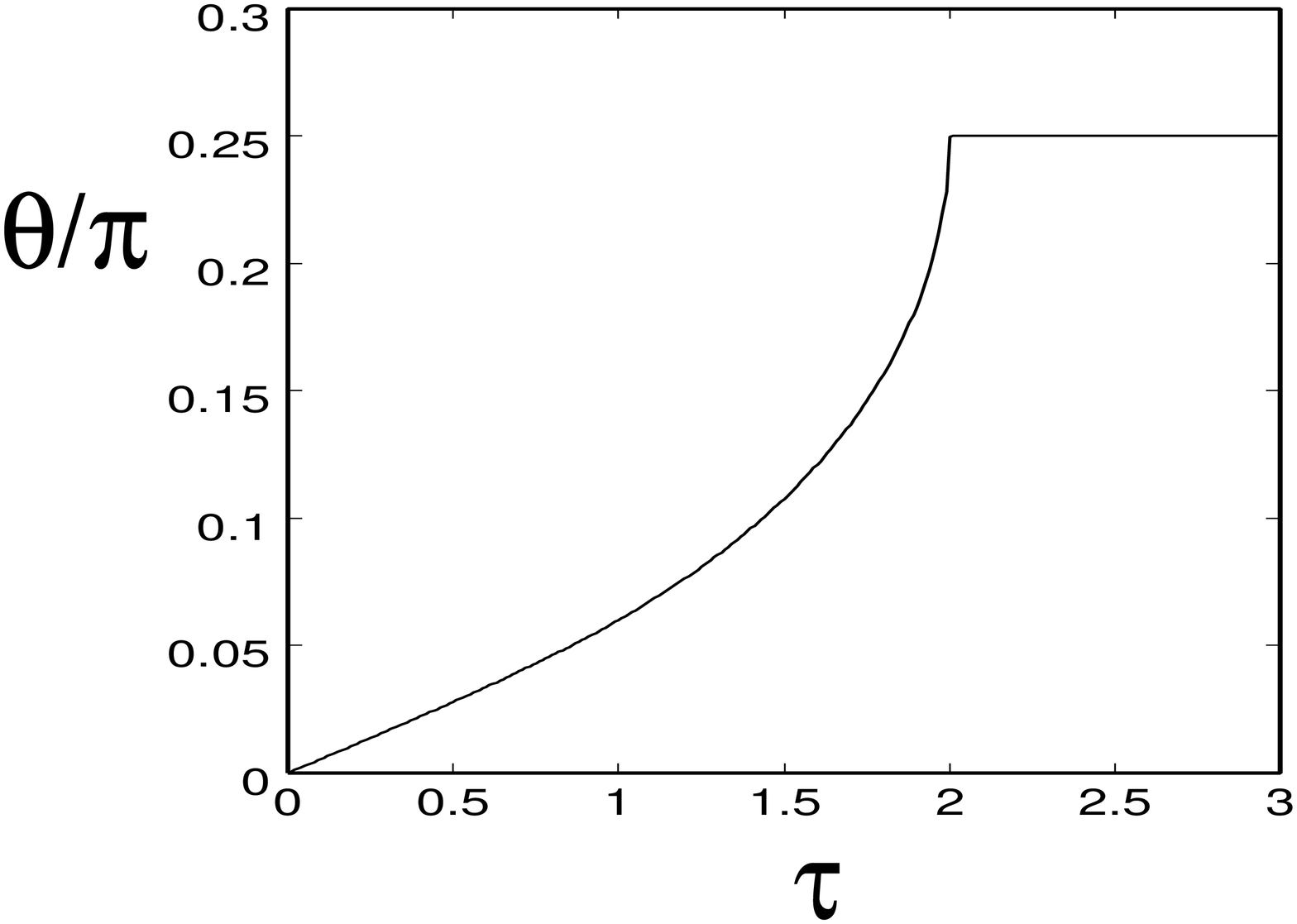,width=2.0in,angle=270}
\vspace{0.1in}
\caption{The inter-layer tunneling dependence of $\theta$.}
\label{tunneling_theta}
\end{figure}

Now let us take into account the inter-layer tunneling effect.
We consider the effect of it on the p-wave pairing state at
$\nu=1/2$. 
The possibility of Haldane-Rezayi state is discussed later.
As we have discussed in Sec.~\ref{sec_two}, the appropriate choice of
$(\phi_1,\phi_2)$ is $(2,2)$.
Since the number of attached fluxes $\phi_1$ and $\phi_2$ is
symmetric, we may take 
$\Delta^{\bf k}_{\uparrow \uparrow} = \Delta_{\bf k} a$ and
$\Delta^{\bf k}_{\uparrow \downarrow}= \Delta_{\bf k} b$.
In order to discuss the evolution of the pairing state,
we define an angle as
\begin{equation}
\theta = \tan^{-1} \frac{a}{b}.
\end{equation}
This angle $\theta$ characterizes the pairing state.
For the case of $\theta=0$, we have the p-wave pairing state that
corresponds to the $(3,3,1)$ state.
Meanwhile, for the case of $\theta=\pi/4$, we have the Pfaffian state.
In Fig.~\ref{tunneling_theta}, we show the inter-layer tunneling
dependence of $\theta$.
Note that the Pfaffian state is stabilized in the $\tau > 2$ region.

\begin{figure}[htbp]
\center
\epsfxsize=2.0truein
\psfig{file=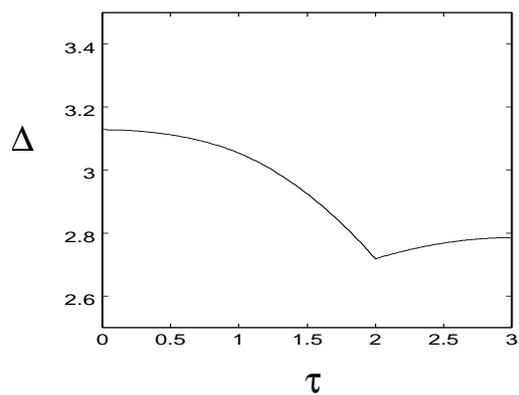,width=2.0in,angle=270}
\vspace{0.1in}
\caption{The inter-layer tunneling dependence of the gap $\Delta$.}
\label{tunneling_gap}
\end{figure}

In Fig.~\ref{tunneling_gap}, we show the inter-layer tunneling
dependence of the gap $\Delta$.
Note that there is a cusp at $\tau=2$.
Reflecting the presence of the cusp in the gap, the ground state
energy also has a cusp at $\tau=2$.

Now we discuss the possibility of the Haldane-Rezayi state, or
d-wave pairing state.
Since the $s$-wave pairing state is excluded as mentioned above, the
next leading singlet pairing state is the d-wave pairing state.
From the analysis of the gap equation, we find that
for the d-wave pairing state $\Delta \sim 1.2$, which is smaller
than that for the p-wave pairing state, in the region of
$\tau < \tau_c$ where $\tau_c \sim 0.8$ and the d-wave pairing
state is not stabilized in $\tau > \tau_c$.
However, in the above analysis we have neglected the effect of the
long-range Coulomb interaction.
Since the effect of it is expected to be larger for the p-wave pairing
state than for the d-wave pairing state, it might be possible that the
d-wave pairing state becomes stable due to the effect of the
long-range Coulomb interaction.
In addition, impurites may affect the p-wave pairing state more than
the d-wave pairing state.

\section{Conclusion}
\label{sec_summary}
In this paper, we have discussed the region of the inter-layer
separation where the p-wave pairing state is stabilized and the
effect of the inter-layer tunneling at $\nu=1/2$.
The Pfaffian state is stable above the critical tunneling strength and 
there is a cusp at the transition point between the Pfaffian state and 
the state continuously connected with the $(3,3,1)$ state.

\acknowledgments
This work was supported in part by a Grant-in-Aid from the Ministry of 
Education, Culture, Sports, Science and Technology.

\end{multicols}

\end{document}